\documentclass[]{aa}
\usepackage{graphicx,amssymb}
\usepackage{natbib}
\bibpunct[, ]{(}{)}{;}{a}{}{,}

\begin{document}
\newcommand{\FF}[1]{{\bf [#1 -- FF.]}}
\newcommand{\avk}[1]{{\bf [#1 -- AVK.]}}
\newcommand{\rev}[1]{{\bf #1}}
\renewcommand{\arraystretch}{1.5}

\title{Planets of $\beta$ Pictoris revisited}

\author{Florian Freistetter \and Alexander V. Krivov \and Torsten L\"ohne}

\offprints{F. Freistetter \email{florian@astro.uni-jena.de}}

\institute{Astrophysikalisches Institut, Friedrich-Schiller-Universit\"at Jena, Schillerg\"asschen 2-3, 07745 Jena, Germany}

\date{Received; accepted}   

\abstract
{
Observations have revealed a large variety of structures (global asymmetries, warps, belts, rings) and 
dynamical phenomena (``falling-evaporating bodies'' or FEBs, the ``$\beta$~Pic dust stream'') in the  
disc of $\beta$ Pictoris, 
most of which may indicate the presence of one or more planets orbiting the star.
Because planets of $\beta$ Pic have not been detected by observations yet, we use dynamical
simulations to find ``numerical evidence'' for a planetary system.
We show that already one planet at 12 AU with a mass of 2 to 5 $M_{\mathrm{J}}$ and an eccentricity
$\la 0.1$ can probably account for three major features (main warp, two inner belts, FEBs) observed in 
the $\beta$ Pic disc.
The existence of at least two additional planets at about 25 AU and 45 AU from the star seems likely.
We find rather strong upper limits of 0.6 $M_{\mathrm{J}}$ and 0.2 $M_{\mathrm{J}}$
on the masses of those planets.
The same planets could, in principle, also account for the outer rings observed at 500 -- 800 AU.


}

\authorrunning{Freistetter et al.}
\titlerunning{Planets of $\beta$ Pictoris revisited}

\maketitle

\section{Introduction}

Since the discovery of the circumstellar disc of $\beta$ Pictoris
by~\citet{smith-terrile-1984}, it became the most observed and best studied debris disc
\citep[see][and references therein]{lagrange-et-al-2000}.
However, a long-standing question whether $\beta$~Pic also hosts
planets, remains unanswered.
Being an A5V star, $\beta$ Pic
is a difficult target for the radial velocity measurements:
the currently achieved precision of hundreds m s$^{-1}$
barely excludes the presence of a $10 M_{\mathrm{J}}$ planet at 1~AU
\citep{galland-et-al-2006}.
The prominent edge-on disc rules out direct imaging.
Transits are not promising either because of their low probability.
At the same time, there is a growing bulk of indirect evidence for the
presence of planets in the system.
\citet{mouillet-et-al-1997} showed that a planet
with an orbital inclination of 3$^\circ$ to 5$^\circ$ and
a mass ranging from 0.6 $M_{\mathrm{J}}$ and 18
$M_{\mathrm{J}}$ between 20 and 3 AU could be responsible for the observed
warp in the disc.
New HST/STIS observations of the warped disc by \citet{heap-et-al-2000}
changed these estimates only slightly.
\citet{augereau-et-al-2001} pointed out that the same planet could explain
the butterfly asymmetry of the disc.
\citet{beust-morbidelli-2000}
found that a jovian planet at $\approx$~10 AU with a moderate
eccentricity of $e\approx 0.05$ can explain both the warp and the observed
phenomenon of falling-evaporating bodies (FEBs).
\citet{krivov-et-al-2004} argued that such a planet is needed to explain the
so-called ``$\beta$~Pic dust stream'' detected by meteor radar AMOR
\citep{baggaley-2000}.
Finally, most recent observations revealed several belt-like structures
in the inner disc (Table~\ref{tab1}) that can be
attributed to the presence of planets. This work is an attempt to
constrain the parameters of suspected planets in the light of
these observations.

\begin{table}
\caption{Observed belt structures in the $\beta$ Pic system}
\label{tab1}
\begin{center}
\begin{tabular}{c c l}
\hline\hline
label & position & ref. \\
\hline 
A & $\approx$ 6.4 AU & 2     \\
B &  $\approx$ 16 AU & 1,2,4 \\
C &  $\approx$ 32 AU & 1,2,4 \\
D &  $\approx$ 52 AU & 1,3,4 \\
\hline
\end{tabular}

\vspace*{1mm}
{\scriptsize
References:\\
1=\citep{wahhaj-et-al-2003},
2=\citep{okamoto-et-al-2004},\\
3=\citep{telesco-et-al-2005},
4=\citep{golimowski-et-al-2006}
}

\end{center}
\end{table}

\section{Presumed planetesimal belts}

\citet{okamoto-et-al-2004}
performed high-resolution spectroscopic observations in the 10-$\mu$m band to identify
concentrations of submicron-sized silicate dust at 6.4 AU, 16 AU and 30 AU and
interpreted these by dust-replenishing planetesimal belts at those locations.
To check this interpretation, we have made test runs of our
collisional code \citep{krivov-et-al-2006}.
The code enables simulations of a circumstellar disc of solids over a wide
range of sizes~-- from planetesimals to fine dust~-- taking into account
stellar gravity, radiation pressure, as well as destructive and cratering collisions.
We took a planetesimal belt of objects with radii from $0.15\mu$m to 7~km and a total mass
of $0.3 M_\oplus$, semimajor axes from 6 and 7~AU and eccentricities between
0.0 and 0.1 and evolved it to a quasi-steady state to obtain a spatial distribution
of dust material sustained by the belt. We then calculated the blackbody thermal
emission of that dust and took a standard line-of-sight integral to obtain
a brightness profile as a function of the projected distance from the star.
The mass and luminosity of $\beta$ Pic were taken to be
$1.75 M_\odot$ and $8 L_\odot$ \citep{crifo-et-al-1997}.
We assumed a bulk density of solids of $2.5$~g~cm$^{-3}$.

\begin{figure}
\begin{center}
\includegraphics[width=9cm]{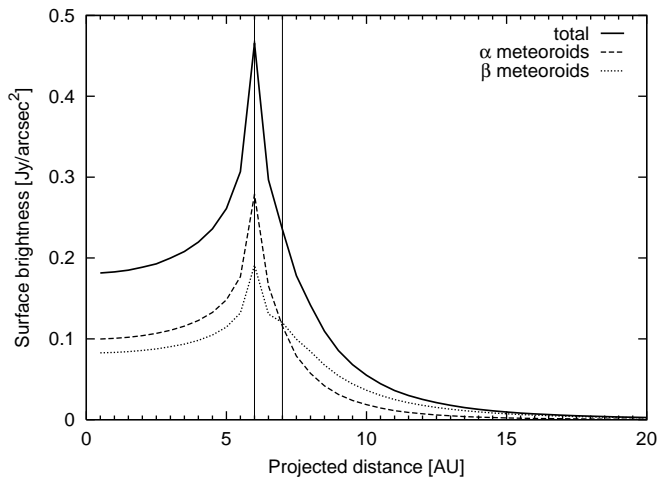}
\caption{The $10\mu$m edge-on brightness of a dust ``subdisc''
produced by a planetesimal belt between 6 and 7~AU
(as shown with vertical lines). Dashed and dotted lines:
contributions from $\alpha-$ and $\beta$-meteoroids, respectively;
solid: their sum.
}

\label{fig1}
\end{center}
\end{figure}

The results are shown in Fig.~\ref{fig1}. The flux is dominated by emission
of grains in bound orbits, or $\alpha$-meteoroids.
However, particles in hyperbolic orbits, or $\beta$-meteoroids,
which are particles smaller than $\approx 2\mu$m,
also make a sensible contribution.
The ``half-peak'' brightness is about a  factor of two larger than in the adjacent parts of
the ring, roughly consistent with the observations
\citep[Fig. 2 of][]{okamoto-et-al-2004}.
A more accurate comparison is not possible, for both a large difference between
the 6.4 AU SW and NE peaks
and the fact that the peak brightness depends on the assumed width of the planetesimal ring.
The total mass of the planetesimal belt of $0.3 M_\oplus$,
comparable to that of the Kuiper belt in the solar system, leads to the peak
brightness $\sim 0.3 \mbox{Jy}\; \mbox{arcsec}^{-2}$, which is close to the
observed values \citep{okamoto-et-al-2004}.
Therefore, we can conclude that the observed brightness peak at $6.4$~AU 
can indeed be attributed to an invisible planetesimal belt at approximately the same location.

The results for the outer belts look similar and therefore are not shown here.
Furthermore, our collisional modelling implies that,
with a reasonable accuracy, possible interaction between the belts can be neglected, and
that they can be treated separately.

\section{Presumed planets}

\subsection{Numerical model of the planetesimal disc}

Numerical simulations of the motion of planetesimals
were carried out with the {\tt mercury6} integration package by
\citet{chambers-1999}.
The planetesimal disc was modeled by a set of massless particles, initially distributed in
equidistant, circular and plain orbits around the star,
in which we placed one or more planets.
Since we want to focus our
study on the inner part of the disc, the orbits of particles ranged from 1 AU to 70 AU with a
step size of $0.1$ AU.
All particles that were in trojan-type motion with one of the planets were removed.
Altogether, we performed about 100 different simulations and in each run
the disc contained 20730 particles.
The integration interval was set to 12 Myr
in accordance with the supposed age of the system
\citep{zuckerman-et-al-2001,ortega-et-al-2002}.

\subsection{One planet}

The first simulations were carried out with one planetary perturber.
\citet{okamoto-et-al-2004} suggested that the borders of the belt at 6.4 AU
are created by resonances with a planet at 12 AU,
similar to the main belt in our solar system which is confined by a 1:2 and 1:4 mean-motion 
resonance (MMR) with  Jupiter. Thus we adopted $a = 12$~AU.
\citet{mouillet-et-al-1997} give a possible range for the inclination of a planet between 
3$^\circ$ and 5$^\circ$ so that we chose $i = 4^\circ$.
Three different values of the eccentricity were tested: $e=0.01$, $e=0.1$ and $e=0.2$.
We tried the planet with several masses between 2 and 5 $M_{\mathrm{J}}$, the values which roughly bracket
the range suggested to explain the warp \citep[see Table 4 of][]{heap-et-al-2000}.

\begin{figure}
\begin{center}
\includegraphics[width=6.3cm,angle=270]{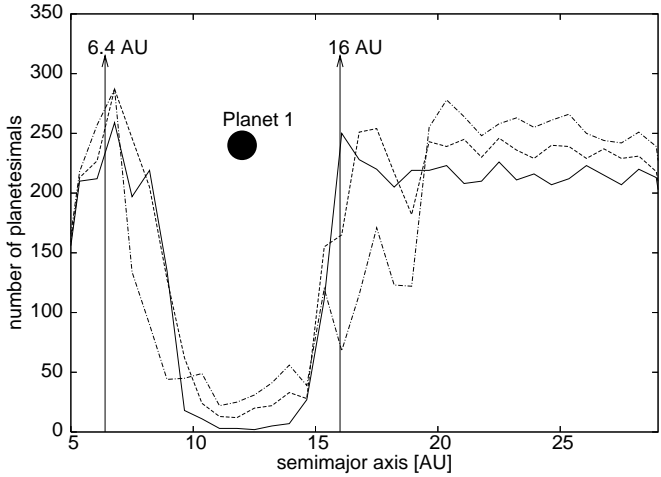}
\caption{Effect of a planet at 12 AU with $m=2M_{\mathrm{J}}$  on 
the inner part of the disc of $\beta$ Pic (solid line: $e=0.01$; dashed line: $e=0.1$; dashed-dotted line: 
$e=0.2$).
}
\label{fig2}
\end{center}
\end{figure}

The simulation results are shown in Fig.~\ref{fig2}.
Note that this and other plots are meant to show the {\em positions} of belts and gaps
only; to determine their strength one would need to convolve the distributions with the assumed radial 
density profile of the disc.
The planet with $m=2M_{\mathrm{J}}$ opens a gap in the disc  
and causes a peak at 6.4 AU and  a smaller peak 
close to 16 AU, corresponding to the A- and B-belts. The eccentricity has no influence on the peak at 
6.4 AU but the peak at 16 AU shifts outwards to $\approx 20$ AU for $e=0.2$. For the planet with 
$m=5M_{\mathrm{J}}$, the peak at 6.4 AU is also present,
but the second peak is at about 20 AU rather than 16 AU
for all three values of eccentricity.
The error bars for the estimated parameters of this planet are listed in Table~\ref{tab2}. 
The limits were determined by changing the semimajor axis,
eccentricity and mass and by checking, whether the resulting peaks in the distribution of test particles still
satisfy the reported observations.
To avoid possible confusion in interpreting the uncertainties of semimajor axes
of the proposed planets, we note that we assumed here the positions of the peaks found by
~\citet{okamoto-et-al-2004} to be exact and give in Table~\ref{tab2} and other places of the paper the
estimated intrinsic error of {\em our} simulations.
However, \citet{okamoto-et-al-2004} have sampled the disc with a step of 3.2~AU,
performing their measurements at discrete distances from the star of 3.2~AU, 6.4~AU, 9.5~AU, and so on.
Therefore, the actual uncertainty of the semimajor axes of the predicted planets is determined
by the uncertainty of the peak locations, $\approx 2$ to $3$~AU.

Although we propose that a second planet around $\beta$ Pic makes the peak near 16 AU 
more prominent (see Sect.~\ref{Sec:23}), already this planet alone could produce a notable peak.
There is an analogy with the {\em Hilda} group of asteroids in our solar system
that move near the 2:3 resonance with Jupiter
and {\em Plutinos} in 3:2 MMR with Neptune \citep[see, e.g.,][]{morbidelli-2002}.
In our case, the 3:2 MMR with the planet at 12 AU is located at $\sim 15.7$ AU.
Many planetesimals in this region are resonant-protected from close encounters with the planet
and form a peak with respect to the non-resonant background population.
Similarly, the inner peak at 6.4 AU is strongly ``supported'' by the 2:5 MMR
in much the same way as the {\em Koronis} asteroid family by its 2:5 MMR with Jupiter.

\begin{figure}
\begin{center}
\includegraphics[width=6.3cm,angle=270]{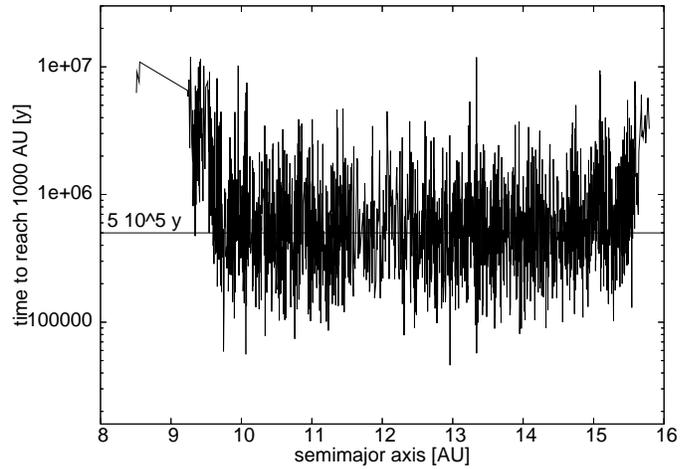}
\caption{Time needed to reach 1000 AU by the scattered particles.}
\label{fig3}
\end{center}
\end{figure}

\begin{figure}
\begin{center}
\includegraphics[width=3.cm,angle=270]{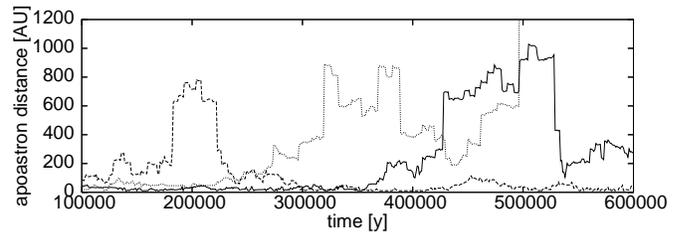}
\caption{Evolution of the apastron distance of three scattered particles}

\label{fig4}
\end{center}
\end{figure}

The particles originally placed near the planet were scattered out of their position;
some of them were thrown out of the system. 
Fig.~\ref{fig3}
shows the time such a particle needs to reach a distance of $10^3$~AU.
The majority need $\sim 0.5$ Myr to reach this distance and thus they might
form some observable features in the outer parts of disc,
for instance the rings observed at 500 -- 800 AU \citep{kalas-et-al-2001}.
Discrete rings may be a result of ejection of several large planetesimals.
Fig.~\ref{fig4} shows, for three of the particles, how close encounters with the 
planet at 12 AU can change successively the apastron distance until it lies 
in the outer part of the disc. 

\citet{beust-morbidelli-2000} analyzed mechanisms proposed to
explain the observed phenomenon of falling-evaporating bodies (FEBs).
All of these~-- close encounters \citep{beust-et-al-1991}, 
Kozai mechanism \citep{bailey-et-al-1992},
trapping in MMRs \citep{beust-morbidelli-1996}~--
involve the presence of at least one large planet.
They showed that a planet with $\approx$ 2 $M_{\mathrm{J}}$ at $\approx$ 10 AU with a low 
eccentricity can account for the detected infall of small bodies onto the star.
Furthermore, they noticed that the parameters of the planet lie well inside the 
limits given by \citet{mouillet-et-al-1997} for a planet that can cause the observed warp.
Because these parameters are close to those we found above,
we can conclude that already a single planet with a mass of 
$m \approx 2M_{\mathrm{J}}$ at 12 AU, $e \la 0.1$ and $i \approx 4^\circ$ can probably 
account for three of the major dynamical phenomena observed in the disc of $\beta$ Pic: the warp,
the A- and B-belts and the FEBs.

\subsection{Two and three planets} 
\label{Sec:23}
\begin{figure}
\begin{center}
\includegraphics[width=6.3cm,angle=270]{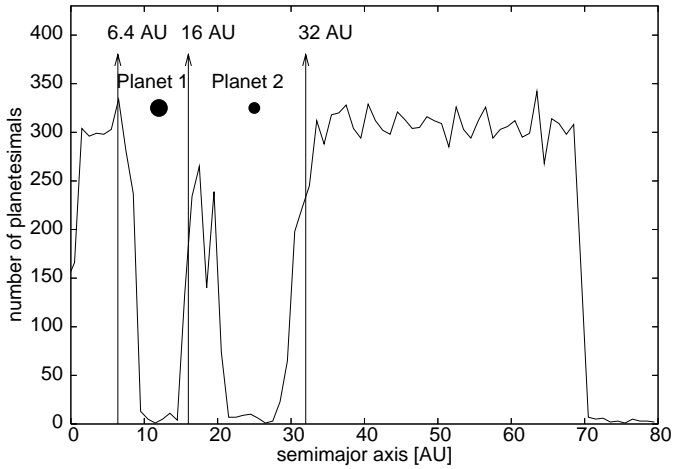}
\caption{Influence of two planets at 12 and 25 AU with  $m=2M_{\mathrm{J}}$ and $m=0.5M_{\mathrm{J}}$ (both with $e=0.01$) on the disc.}

\label{fig5}
\end{center} 
\end{figure}

The second planet can be invoked to account for the peak around 32 AU (C-belt) that could not be 
created by the planet at 12 AU.
If we  follow the same strategy as for the first planet, we find that the best fit is a planet with 
$m=0.5M_{\mathrm{J}}$ at $a=25$ AU and $e=0.01$ that clears a gap and causes a new peak at 32 AU. 
The distribution of particles in the two-planet case
is shown in Fig.~\ref{fig5}.

\begin{figure}
\begin{center}
\includegraphics[width=6.3cm,angle=270]{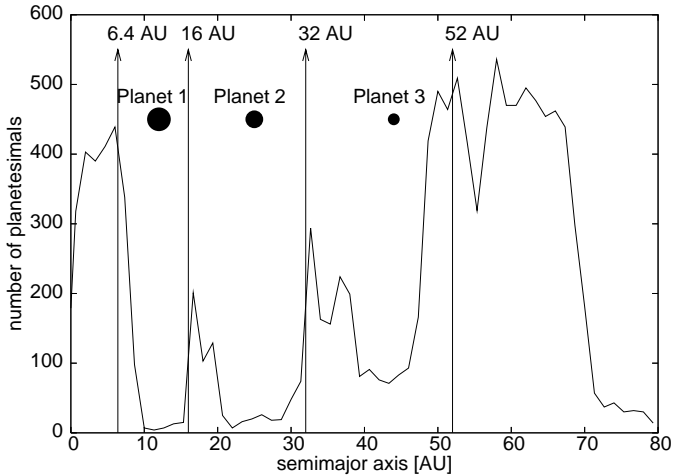}
\caption{Effect of three planets at 12, 25 and 44 AU with $m=2M_{\mathrm{J}}$,  $0.5M_{\mathrm{J}}$ and  
$0.1M_{\mathrm{J}}$ (all with $e=0.01$) on the disc.}
\label{fig6}
\end{center}
\end{figure}

Similarly, a third planet is needed to explain the peak around 52 AU (D-belt).
The best fit is with $m=0.1M_{\mathrm{J}}$ at 44~AU.
Figure~\ref{fig6} 
shows the result of our simulations with three planets.
The D-belt is now created, and
the peak at 32 AU became more prominent because it is now bordered by 
a planet on both sides.
Also, the peaks at 6.4 and 16 AU are still present. 
It is important to mention that a more massive second planet, with
a mass larger than $\approx 0.6 M_{\mathrm{J}}$,
can be excluded because it would destroy the belts.
The same applies to the third planet which cannot be 
more massive than $\approx 0.2M_{\mathrm{J}}$.
Table~\ref{tab2} summarizes the best-fit parameters of all three planets
and their errors bars. 

\begin{table}
\caption{Parameters for the proposed $\beta$ Pic planetary system}
\label{tab2}
\begin{center}
\begin{tabular}{c l l l}
\hline\hline
Planet & $m$ [$M_{\mathrm{J}}$] & $a$ [AU] & $e$\\
\hline 
1 & $2.0^{+3}_{-0.5}$     & 12 $\pm$ 0.5 & $0.01^{+0.1}_{-0.01}$\\
2 & $0.5 \pm 0.1$         & 25 $\pm$ 1   & $0.01^{+0.05}_{-0.01}$\\
3 & $0.1^{+0.1}_{-0.03}$  & 44 $\pm$ 1   & $0.01^{+0.05}_{-0.01}$\\
\hline
\end{tabular}

\end{center}
\end{table}

\section{$\beta$ Pic -- a resonant system?}

The orbital periods of the proposed three planets are close to rational commensurabilities.
Namely, the periods of planet 2 and planet 1 are nearly in a 3:1 ratio;
those of planet 3 and planet 1 are close to a 7:1 ratio; and planet 3 and
planet 2 are near a 7:3 commensurability.
Planets in some of the extrasolar planetary systems discovered so far are
known to be locked in mean-motion resonances (e.g. GJ876 or 55 Cnc). To see
if the possible planets of $\beta$ Pic really show resonant motion,
we have chosen the strongest of the three resonances mentioned above,
the 3:1 MMR between the two inner planets, and calculated the
resonant angle \citep{ji-et-al-2003}
\begin{equation}
\theta=\lambda_1 - 3\lambda_2 + \left(\tilde{{\omega_1}} + \tilde{{\omega_2}}\right),
\end{equation}
where $\lambda_i$ and $\tilde{{\omega_i}}$ are the true longitude and the
longitude of periastron of the $i$-th planet.

\begin{figure*}
\begin{center}
\includegraphics[width=6cm,angle=270]{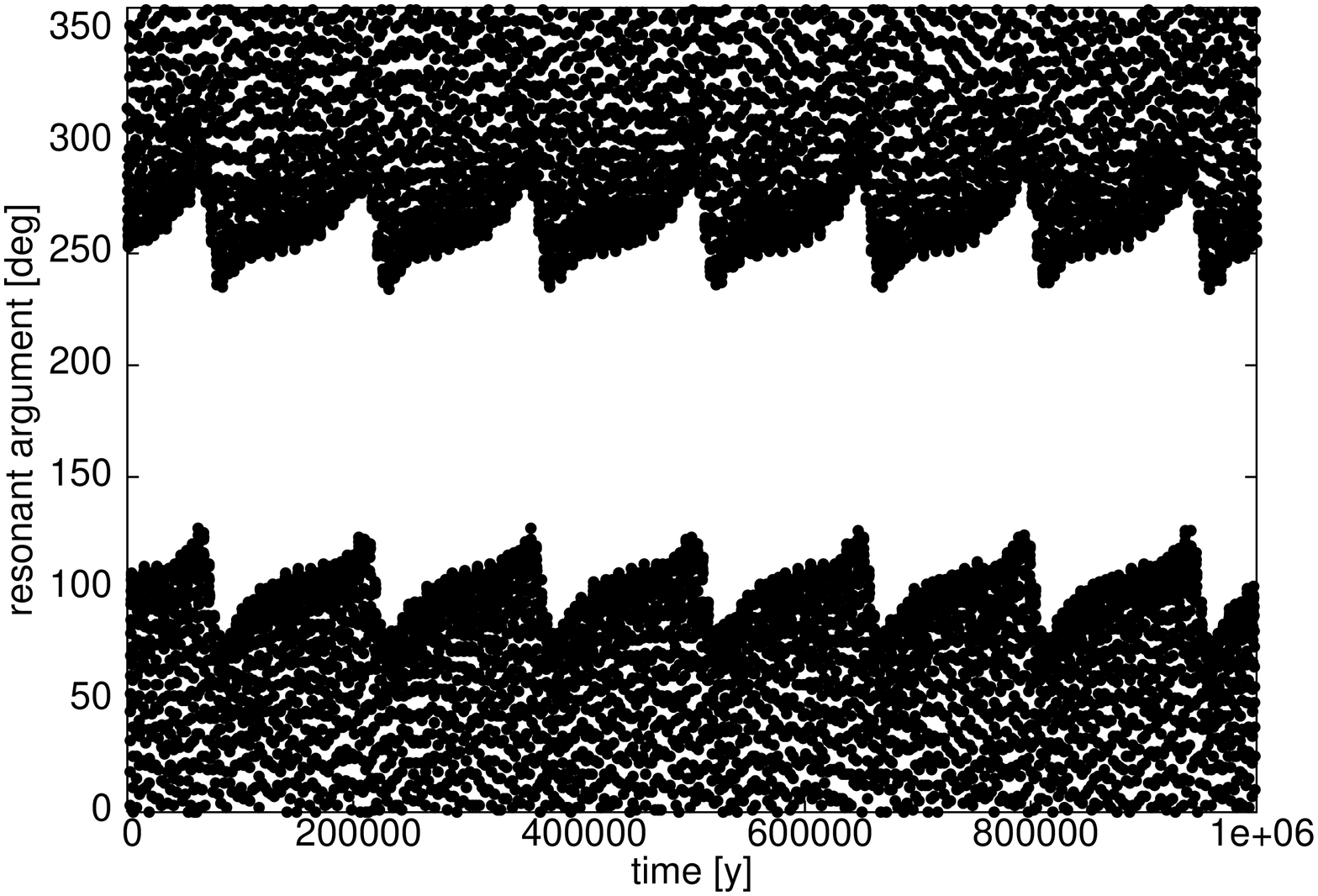}
\includegraphics[width=6cm,angle=270]{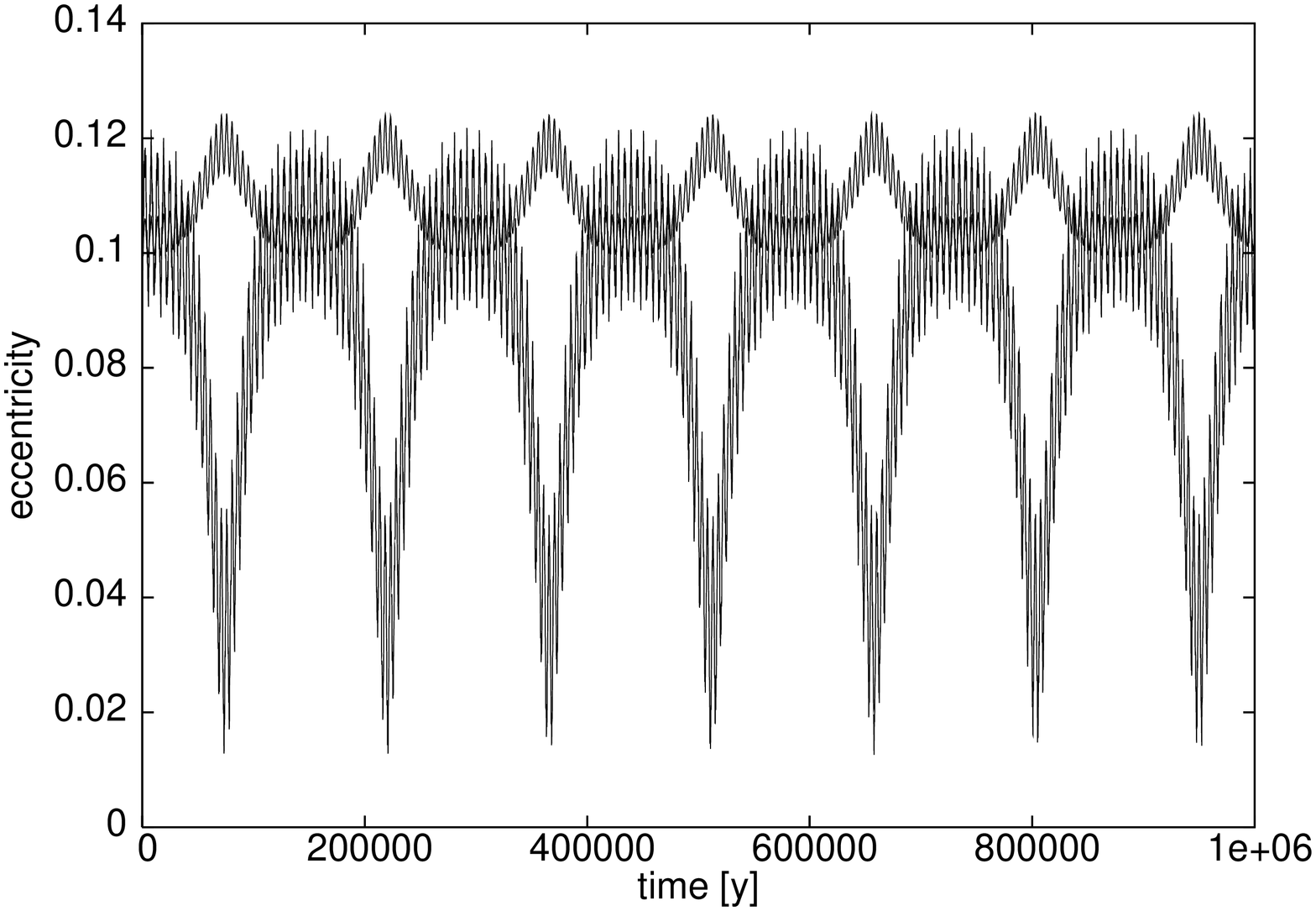}
\includegraphics[width=6cm,angle=270]{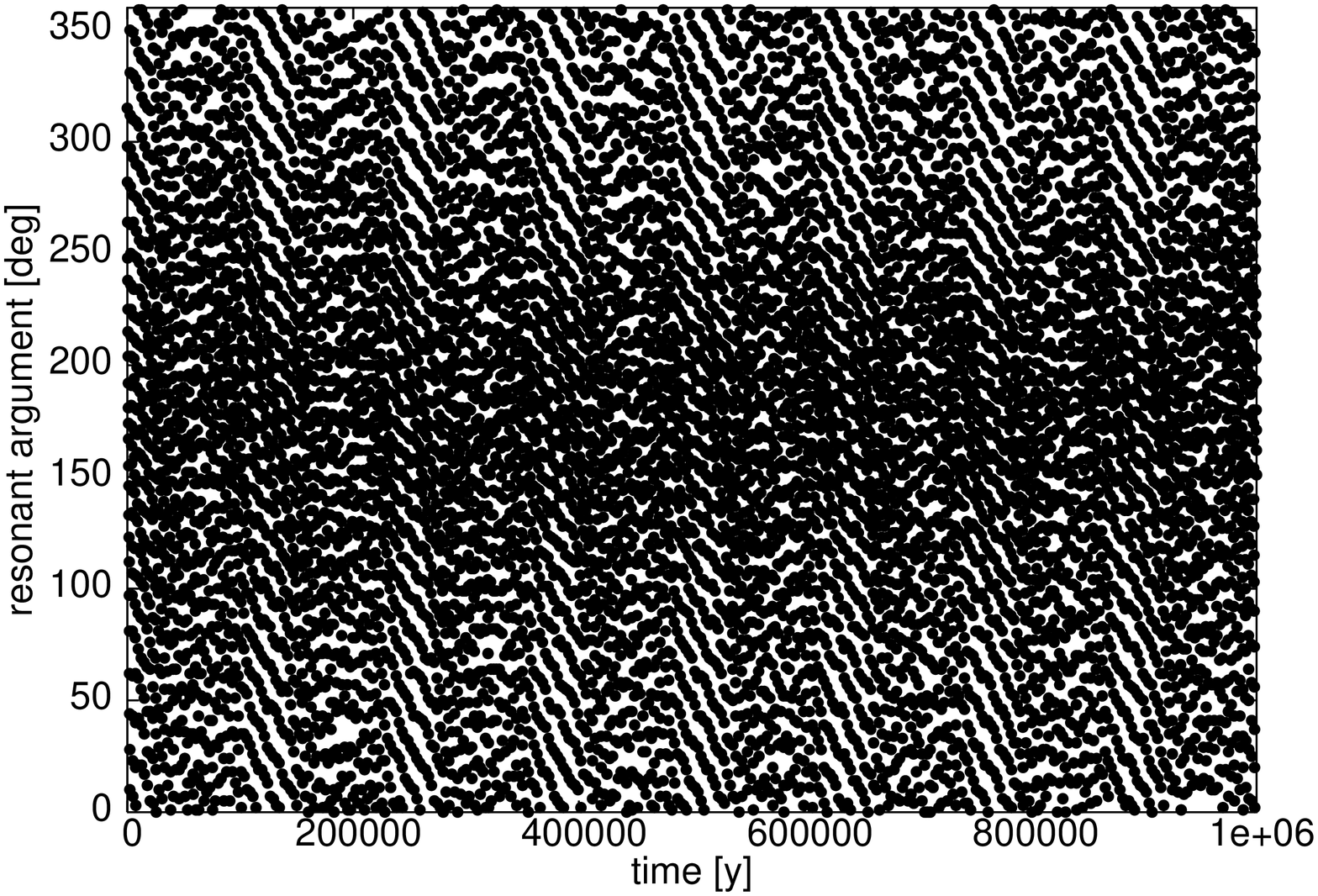}
\includegraphics[width=6cm,angle=270]{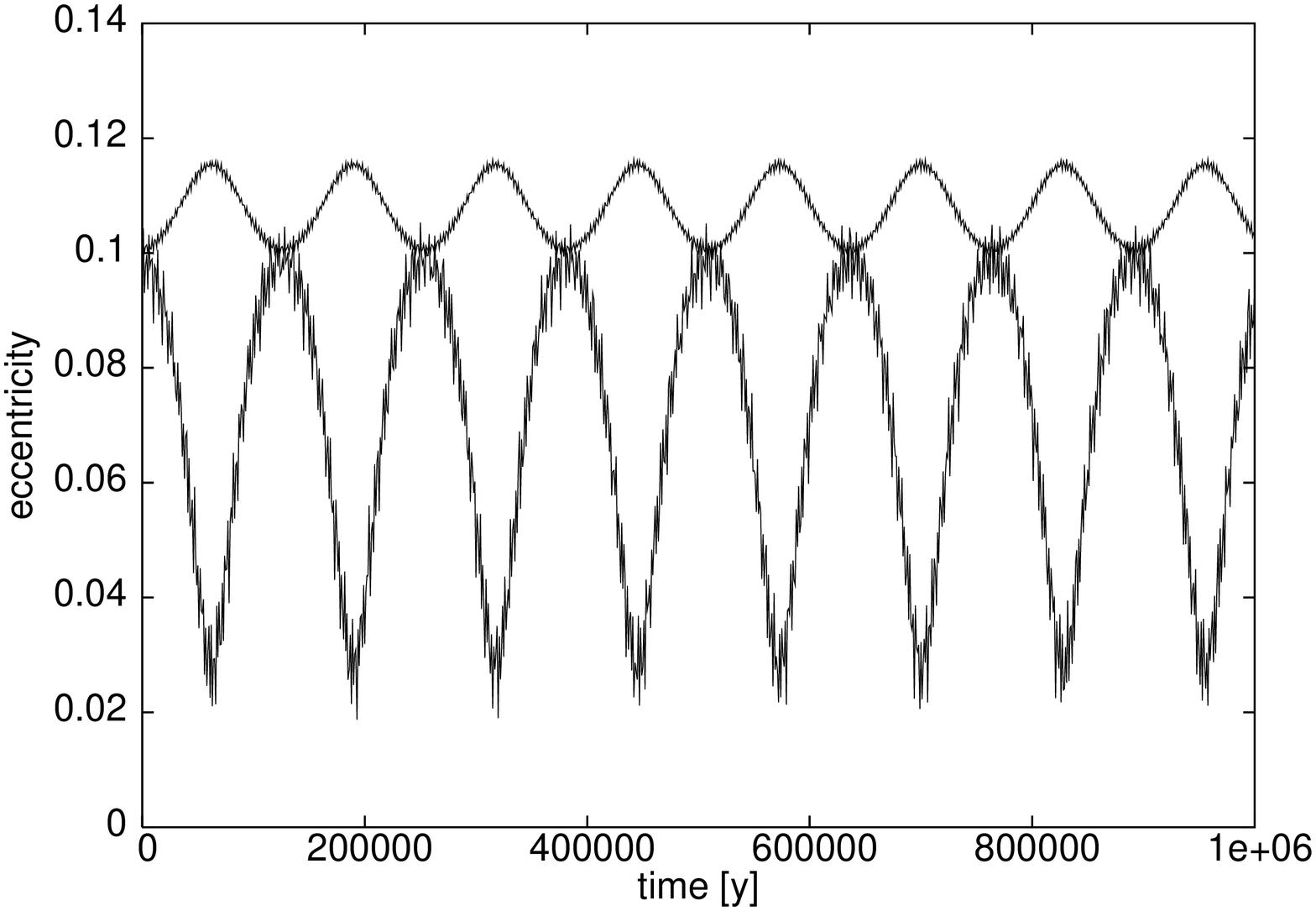}
\caption{Resonant angle $\theta$ for the 3:1 resonance of planet 1 and planet 2 (left) and evolution of the 
eccentricities (right). Top: the case were planet 2 is located at the exact position of 
the resonance. Bottom: planet 2 is shifted by $0.1$~AU inwards. Both planets 
have an initial eccentricity of $0.1$ and an inclination of $4^\circ$.}
\label{fig7}
\end{center}
\end{figure*}

Figure \ref{fig7} (top) shows the case, where planet 2 is located at the exact position of the resonance
($a_1 = 12.00{\rm AU}$ and $a_2 = 24.96{\rm AU}$); both 
planets have an initial eccentricity of $0.1$.
Other parameters are the same as in the simulations described in Sec.~\ref{Sec:23}.
The resonant angle librates around $0^\circ$, indicating the resonant locking.
However, the eccentricties are not strongly affected by the resonance which seems to be a 
protective one, as it is the case for GJ876 and 55 Cnc.
That the effect is weak is not surprizing: the resonance is rather shallow (as seen from a rather
large libration amplitude), which, in turn, traces back to moderate masses of both planets
and a large separation between their orbits.

We have found that a decrease in semimajor axis of the second planet
by $\approx 0.1$~AU or an increase by $\approx 0.2$~AU
is enough to transfer $\theta$ from libration to a circulation mode 
(Fig.~\ref{fig7}, bottom). 
This value can be compared to analytic estimates of the resonance width.
The semimajor axis of a body locked in a resonance oscillates between
$a - \Delta a_\mathrm{max}$ and $a + \Delta a_\mathrm{max}$, where
the libration amplitude is given by
\citep{murray-dermott-1999}
\begin{equation}
  \Delta a_\mathrm{max} =
    a_2 \left[ \frac{16}{3} \frac{m_\mathrm{1}}{m_*} \left( \frac{p}{p+q} \right) ^ {\frac 2 3} f_\mathrm{d} e_2 \right] ^ {\frac 1 2}
\label{eq1}
\end{equation}
with $f_\mathrm{d} = 0.5988$
for the $(p+q):p=3:1$ outer resonance.
Assuming $e_2 = 0.1$,
this leads to
\begin{equation}
  \Delta a_\mathrm{max} \approx 0.3 \rm{AU} ,
\end{equation}
which is close to the value we found empirically.
Thus the resonance width is smaller than
the uncertainties in the semimajor axes of the planets
($\Delta a \sim 1 AU$ for our fits; $\Delta a \sim 3 AU$ for the original
observations of \citet{okamoto-et-al-2004}).

The eccentricity plays an important role as well. Our
best fits for the three planets imply small initial
eccentricities ($ e\sim 0.01$). With these small values, the resonant argument
in our simulations was always circulating rather than librating.
Libration of the resonant angle was observed, starting from $e \ga 0.07$.
The fact that the resonance gets thinner at low eccentricities is also seen
from Eq.~\ref{eq1}: $\Delta a_\mathrm{max} \propto \sqrt{e_2}$.
Since the uncertainties of the derived semimajor axes are much larger than the precision required to 
distinguish resonant from non-resonant configurations, we cannot conclude whether the possible $\beta$~Pic 
system is a resonant one or not. As mentioned above, some
exoplanetary systems are known to be resonant. It is possible that a planetary system
emerges as a resonant system~--- for instance, as a result of differential migration
in a gaseous disc \citep[e.g.,][]{ferrazmello-et-al-2005,kley-et-al-2005}. Alternatively, the $\beta$~Pic 
system could resemble the Jupiter-Saturn configuration in our Solar System. Here the two planets are 
close to, but not locked in, a $5:2$ resonance.

\section{Conclusions}

We have used numerical simulations to investigate the effect of one or more planets on the disc of 
$\beta$ Pic. The goal was to find a minimum set of planetary perturbers which
could be responsible for as many features/phenomena observed in the disc
as possible.

\begin{figure}
\begin{center}
\includegraphics[width=8.5cm,angle=0]{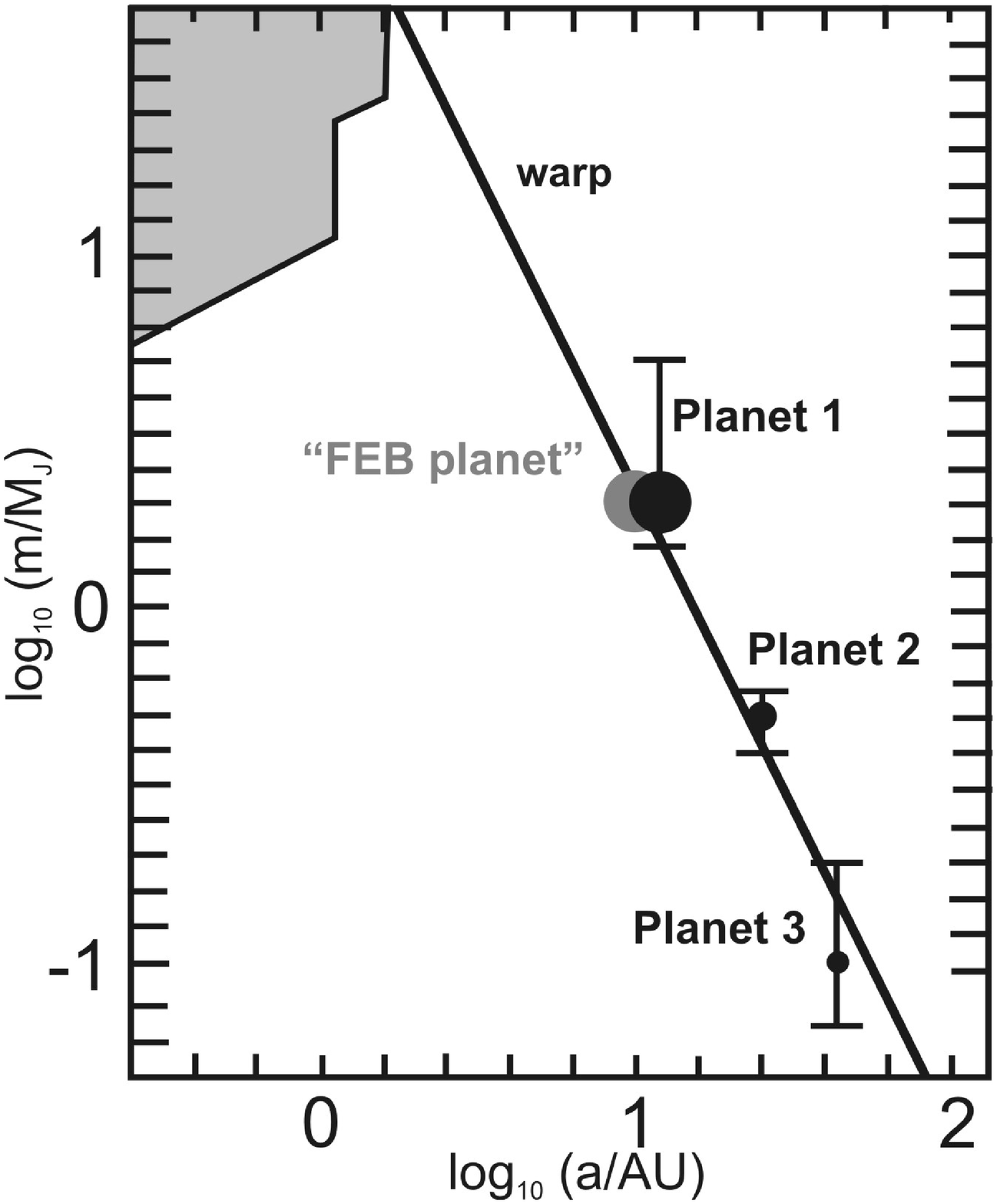}
\caption{$\log a$--$\log m$ phase space for alleged $\beta$ Pic planet(s).
The regions where radial velocity studies exclude a planet are shown in grey
\citep{galland-et-al-2006}.
Line: a planet that could cause the observed warp 
\citep{mouillet-et-al-1997}.
Grey circle: the ``FEB-planet''of \citet{beust-morbidelli-2000}.
The positions of three planets suggested by this study are shown with black circles
with vertical error bars reflecting the uncertainty in the planet masses.
Horizontal error bars would be indistinguishably small and therefore are not shown.
}
\label{fig8}
\end{center}
\end{figure}

One important result is that already one planet at $\approx$ 12 AU with a mass of 
$m \approx 2M_{\mathrm{J}}$ and an eccentricity of $e \la 0.1$ is able to cause three major known features:
(i) two of the four
belt-like structures listed in Table~\ref{tab1};
(ii) the parameters of the planet lie well inside the 
limits given by \citet{mouillet-et-al-1997} for a planet responsible to the warp; and
(iii) this planet is similar to the one proposed
by \citet{beust-morbidelli-2000} to explain the FEB phenomenon 
($a \approx$ 10 AU, $m=2M_{\mathrm{J}}$, low $e$),
so that the two can be considered identical.
Another result is that two additional, more distant planets
would naturally explain further planetesimal belts suggested by some observations.
We find rather strong upper limits on the masses of those planets:
$\approx 2M_{\mathrm{SATURN}}$ for a second planet at 25 AU and
$\approx 4M_{\mathrm{NEPTUNE}}$ for a third one at 44 AU.
More massive perturbers would destroy the belts.
Figure~\ref{fig8} shows
the positions of the proposed planets in the semimajor axis--mass plane.
Overplotted are regions excluded by radial velocity measurements \citep{galland-et-al-2006}, as well as 
the phase-space locations of the warp-inducing planet and the ``FEB planet''.
Interestingly, all three planets are close to the ``warp planet line''.

The same planet(s) could, in principle, also account for 
the outer rings observed at 500 -- 800 AU from the star \citep{kalas-et-al-2001}.
A few large planetesimals could have encountered one of the inner planets in the past and,
perhaps after tidal disruption during these encounters, been sent by the planets into
escaping orbits or those with apocenters at hundreds of AU where the rings are observed.
Further work is also needed to answer the question if the planetary system of $\beta$ Pic is a resonant one or 
not. If it really exists and the planets are in a resonant configuration, this may indicate that migration 
played an important role in its formation.

\acknowledgements
{We wish to thank Philippe Th\'ebault, Jean-Charles Augereau, Doug Hamilton,
Rudolf Dvorak, Sylvio Ferraz-Mello and Elke Pilat-Lohinger
for stimulating discussions and comments on the manuscript.
A speedy and helpful review provided by Herv{\'e} Beust is appreciated.}

\newcommand{\AAp}      {\aap}
\newcommand{\AJ}       {\aj}
\newcommand{\ApJ}      {\apj}
\newcommand{\JGR}      {\jgr}
\newcommand{\MNRAS}     {\mnras}
\bibliography{english}
\bibliographystyle{aa}

\end{document}